\documentclass[twocolumn,showpacs,preprintnumbers,amsmath,amssymb,
superscriptaddress,prl,floatfix]{revtex4}
\usepackage{mathrsfs}
\usepackage{longtable}
\usepackage{graphicx}
\usepackage{epsfig}
\usepackage{dcolumn}

\begin{document}

\title{Cross-Section Measurements of the $^{86}$Kr$(\gamma,n)$ Reaction to Probe the $s$-Process
Branching at $^{85}$Kr}

\author{R.\,Raut}
\altaffiliation[Present Address:]{UGC-DAE Consortium for Scientific Research, Kolkata Centre, Kolkata, India}
\affiliation{Department of Physics, Duke University, Durham, North Carolina 27708, USA}
\affiliation{Triangle Universities Nuclear Laboratory, Durham, North Carolina 27708, USA}
\author{A.\,P.\,Tonchev}
\altaffiliation[Present Address:]{Lawrence Livermore National Laboratory, Livermore, California, USA}
\affiliation{Department of Physics, Duke University, Durham, North Carolina 27708, USA}
\affiliation{Triangle Universities Nuclear Laboratory, Durham, North Carolina 27708, USA}
\author{G.\,Rusev}
\altaffiliation[Present Address:]{Los Alamos National Laboratory, Los Alamos, New Mexico, USA}
\affiliation{Department of Physics, Duke University, Durham, North Carolina 27708, USA}
\affiliation{Triangle Universities Nuclear Laboratory, Durham, North Carolina 27708, USA}
\author{W.\,Tornow}
\affiliation{Department of Physics, Duke University, Durham, North Carolina 27708, USA}
\affiliation{Triangle Universities Nuclear Laboratory, Durham, North Carolina 27708, USA}
\author{C.\,Iliadis}
\affiliation{Department of Physics \& Astronomy, University of North Carolina at Chapel Hill, Chapel Hill, North Carolina 27695-8202, USA}
\affiliation{Triangle Universities Nuclear Laboratory, Durham, North Carolina 27708, USA}
\author{M.\,Lugaro}
\affiliation{Monash Centre for Astrophysics (MoCA), Monash University, Victoria 3800, Australia}
\author{J.\,Buntain}
\affiliation{Monash Centre for Astrophysics (MoCA), Monash University, Victoria 3800, Australia}
\author{S.\,Goriely}
\affiliation{Institut d' Astronomie et d' Astrophysique, Universit\'{e} Libre de Bruxelles, CP 226, 1050 Brussels, Belgium}
\author{J.\,H.\,Kelley}
\affiliation{Triangle Universities Nuclear Laboratory, Durham, North Carolina 27708, USA}
\affiliation{Department of Physics, North Carolina State University, Raleigh, North Carolina 27695, USA}
\author{R.\,Schwengner}
\affiliation{Institut f\"ur Strahlenphysik, Helmholtz-Zentrum Dresden-Rossendorf, 01314 Dresden, Germany}
\author{A.\,Banu}
\affiliation{Department of Physics and Astronomy, James Madison University, Harrisonburg, Virginia 22807, USA}
\author{N.\,Tsoneva}
\affiliation{Institut f\"{u}r Theoretische Physik, Universit\"{a}t Gie$\beta$en, Gie$\beta$en, D-35392, Germany}
\affiliation{Institute of Nuclear Research and Nuclear Energy, 1784 Sofia, Bulgaria}

\date{\today}

\begin{abstract}

We have carried out photodisintegration cross-section measurements on $^{86}$Kr using  monoenergetic photon beams ranging from the neutron separation energy, $S_n = 9.86$ MeV, to 13 MeV. We combine our experimental $^{86}$Kr($\gamma$,n)$^{85}$Kr cross section with results from our recent $^{86}$Kr($\gamma$,$\gamma^\prime$) measurement below the neutron separation energy to obtain the complete nuclear dipole response of $^{86}$Kr.  The new experimental information is used to predict the neutron capture cross section of $^{85}$Kr, an important branching point nucleus on the abundance flow path during $s$-process nucleosynthesis. Our new and more precise $^{85}$Kr(n,$\gamma$)$^{86}$Kr cross section allows to produce more precise predictions of the $^{86}$Kr abundance from $s$-process models. In particular, we find that the models of the $s$-process in asymptotic giant branch stars of mass $< 1.5 M_\odot$, where the $^{13}$C neutron source burns convectively rather than radiatively, represent a possible solution for the highest $^{86}$Kr/$^{82}$Kr ratios observed in meteoritic stardust SiC grains.
\end{abstract}

\maketitle

Stars with masses smaller than about eight solar masses, $\lesssim$8~M$_{\odot}$, become asymptotic giant branch (AGB) stars during their final evolutionary stage. Such stars experience episodical thermal pulses in the helium-burning shell, when the entire helium-rich layer becomes convective for a few hundred years. After each thermal pulse, the convective envelope can reach into the helium-rich layer and dredge-up the products of nucleosynthesis to the stellar surface, from where strong stellar winds expel them into the interstellar medium \cite{Her05,Ili06}. Asymptotic giant branch stars are predicted to be the source of about half of all elements beyond iron in the Galaxy \cite{Bus99}. These elements are produced  in AGB stars via {\it slow} neutron capture (the $s$-process), which operates at relatively low neutron densities. Under such conditions most short-lived radioactive nuclei reached by the $s$-process undergo $\beta$-decay rather than neutron capture. When the $s$-process reaches longer-lived radioactive nuclei, however, neutron capture may compete with $\beta$-decay, giving rise to $s$-process {\it branchings}.\\ 

\begin{figure}
\includegraphics[angle=0,scale=.3]{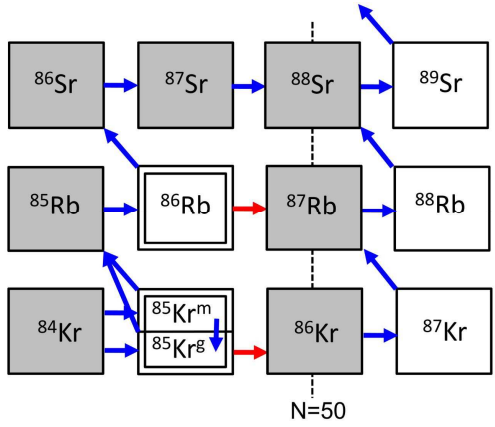}
\caption{\label{fig-chart} 
Nuclidic chart near the $s$-process branching at $^{85}$Kr. Nuclides shown as shaded (open) squares are stable (radioactive). The blue (red) arrows depict the $s$-process path at low (high) neutron densities. The ground state of $^{85}$Kr can either $\beta$-decay ($T_{1/2}=10.75$~y) to $^{85}$Rb or capture a neutron to form $^{86}$Kr. The isomeric state of $^{85}$Kr can either $\beta$-decay ($T_{1/2}=4.48$~h) or decay to the ground state via $\gamma$-ray emission; its neutron capture rate is negligible in $s$-process models. Another s-process branching occurs at $^{86}$Rb. The dashed vertical line connects nuclides with a magic number of neutrons ($N=50$).} 
\end{figure}

The important branching point at $^{85}$Kr involves the additional complication of an isomeric state that is populated via neutron capture on $^{84}$Kr. The situation is depicted in Fig.~\ref{fig-chart}. When the stable $^{84}$Kr nucleus captures a neutron, the excited $^{85}$Kr compound nucleus can either de-excite to the ground state, $^{85}$Kr$^g$, or to the isomeric level, $^{85}$Kr$^m$, with roughly the same probability \cite{Bee91}. The ground state $\beta$-decays with a half-life of $T_{1/2}=10.75$~y to $^{85}$Rb, while the isomeric level can either $\beta$-decay ($T_{1/2}=4.48$~h; $\approx$80\% probability) or $\gamma$-ray decay ($\approx$20\% probability) to the ground state. Because of the long half-life, the ground state of $^{85}$Kr preferably captures a neutron when the neutron density exceeds $5\times10^{9}/cm^3$, and thus becomes a bridge for the production of $^{86}$Kr at elevated neutron densities. Consequently, the branching at $^{85}$Kr is well suited for an estimate of the s-process neutron density. Possible thermalization of the isomer could reduce the effective half-life of $^{85}$Kr ground state, however, this effect appears to be relevant only for temperatures higher than 300 MK \cite{Tak87}, above those of typical AGB $s$-process conditions.\\

The abundance ratio of $^{86}$Kr to any stable krypton isotope on the main $s$-process path, such as $^{82}$Kr, is strongly influenced by the operation of the $^{85}$Kr branching. Precise $^{86}$Kr/$^{82}$Kr ratios have been derived from measurements of krypton atoms trapped inside stardust silicon carbide (SiC) grains \cite{Lew94}, which formed in carbon-rich AGB stars and are recovered from primitive meteorites. The krypton and other noble gas atoms were presumably implanted, after ionization, in already formed SiC grains \cite{Ver04}. However, it is difficult for current stellar models to account for the highest observed $^{86}$Kr/$^{82}$Kr ratios up to $\simeq$ 3 \cite{Pign06}. An outstanding problem is that the predicted $^{86}$Kr/$^{82}$Kr ratios depend sensitively on the $^{85}$Kr($n,\gamma$)$^{86}$Kr reaction rate \cite{Lug05} and that this particular rate is currently rather uncertain. Since $^{85}$Kr is radioactive, the $^{85}$Kr($n,\gamma$)$^{86}$Kr reaction has not been measured directly so far. There is only one measurement by Bemis {\it{et al.}} \cite{Bem71} at thermal neutron energies, presumably dominated by isolated resonances, with no straightforward way to extrapolate the measured cross section to the keV regime of relevance in $s$-process studies. Theoretical estimates of the $^{85}$Kr($n,\gamma$)$^{86}$Kr reaction rate are compiled in Ref.~\cite{Kadonis}. For example, at the energy of $kT=30$ keV ($T=350$ MK), traditionally used in discussions of the $s$-process, the reported Maxwellian-averaged cross-section values range between 25~mb and 150~mb. Most modern $s$-process calculations adopt the value of 55 $\pm$ 45 mb recommended by Bao et al. \cite{bao00}. Clearly, an accurate estimate of $^{85}$Kr($n,\gamma$)$^{86}$Kr cross section in the relevant energy range is essential for extracting meaningful information on the {\it{s}}-process from the investigation of the $^{85}$Kr branching point.\\ 

\begin{figure}
\vspace{-10mm}
\includegraphics[angle=0,scale=.35]{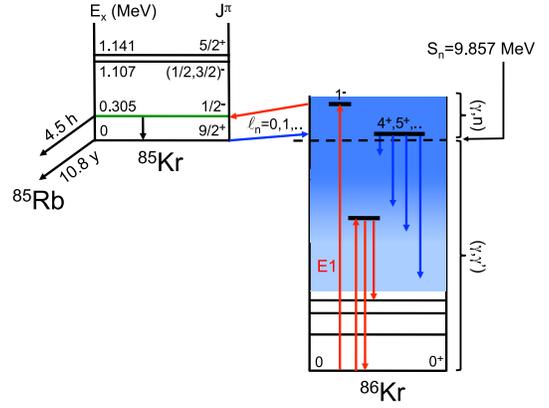}
\caption{\label{fig-scheme}
(Color online) Level schemes of $^{85}$Kr and $^{86}$Kr. The blue arrows depict the ($n,\gamma$) reaction, while the red arrows correspond to the ($\gamma$,n) and ($\gamma$,$\gamma^\prime$) reactions above and below the neutron threshold (thick dashed line), respectively. The blue-shaded region depicts the increasing level density at higher excitation energies. See text.
}
\end{figure}

The present work reports the cross-section measurements of the $^{86}$Kr($\gamma,n$)$^{85}$Kr photoneutron reaction at energies from near threshold (S$_n$ = 9.857 MeV, for $^{86}$Kr) to 13.0 MeV. Initial results at a few of these energies have been previously published by our group \cite{Rau12}. The impetus is to optimize the nuclear ingredients that reproduce the ($\gamma,n$) and ($\gamma,\gamma^\prime$) cross sections and to apply the same input for the reverse ($n,\gamma$) reaction to further constrain the neutron-capture cross section for the $^{85}$Kr branching point nucleus. The mechanisms involving $(\gamma,n)$ and $(\gamma,\gamma^\prime)$ processes are illustrated in Fig.~\ref{fig-scheme}. At very low excitation energies such as $E_\gamma < S_n+305~keV$, the $^{86}$Kr($\gamma,n$) reaction can proceed only to the ground state of $^{85}$Kr. It should be noted that this path would be strongly hampered due to the large angular momentum required for the emitted neutrons ($f-wave$) from the compound nucleus $^{86}$Kr$^*$ with $J = 1$ to the ground state of $^{85}$Kr with $J^\pi = 9/2^+$. At higher energies, population of the ground state proceeds predominantly via the population of the excited states in $^{85}$Kr which then $\gamma$-decay to the ground state.\\ 

The present measurements were carried out at the High Intensity Gamma-Ray Source (HI$\gamma$S) facility of the Triangle Universities Nuclear Laboratory (TUNL). HI$\gamma$S is the most intense source of monoenergetic photon beams in the world, with an average flux of $\sim$ 10$^7$-10$^8$ $\gamma$/s in the energy range from 1 to 20 MeV and typical energy spread of 1-3\%. The target consisted of 1012 mg of Kr gas enriched to 99.4\% in $^{86}$Kr, contained in a stainless steel cell. An empty cell of identical material and dimension was used to subtract the background contribution. The emitted neutrons from the $(\gamma,n)$ reaction were detected using a 4$\pi$ assembly of $^3$He proportional counters, fabricated into a single unit. The efficiency of the detector was extensively studied previously \cite{Arn11} and, in the present work, this detector was operated under identical conditions as described therein. The photon flux incident on the target was measured using a thin plastic scintillator that was cross-calibrated against the $^{197}$Au($\gamma,n$) reaction \cite{Vog02} between 10.0 and 13.0 MeV. For this purpose, very thin Au-foils with the same diameter as the photon beam diameter of 1.905 cm and thickness of 50 $\mu$m were positioned at the exit end of collimator \cite{Ton10}.  The induced activity of $^{196}$Au$^g$ was measured off-line using a well calibrated HPGe detector. The photoneutron cross section was determined from the formula
\begin{equation}
\sigma(E_\gamma) = \frac{N_n}{N_\gamma N_t \epsilon_n g},
\end{equation}
where $N_n$ is the net number of neutrons, $N_\gamma$ is the number of incident photons, $N_t$ is the number of target atoms, $\epsilon_n$ is the neutron detection efficiency and $g$ is the fraction of the incident $\gamma$-rays with energy higher than the neutron separation energy of $^{86}$Kr. In addition, at four incident photon energies the Kr gas cell was removed from the bore of $^3$He proportional counter and the induced activity of $^{85}$Kr$^m$ (T$_{1/2}$ = 4.48 h) was measured with the same HPGe detector as the gold activation foils.\\ 

\begin{figure}
\includegraphics[angle=0,scale=.30,clip=true]{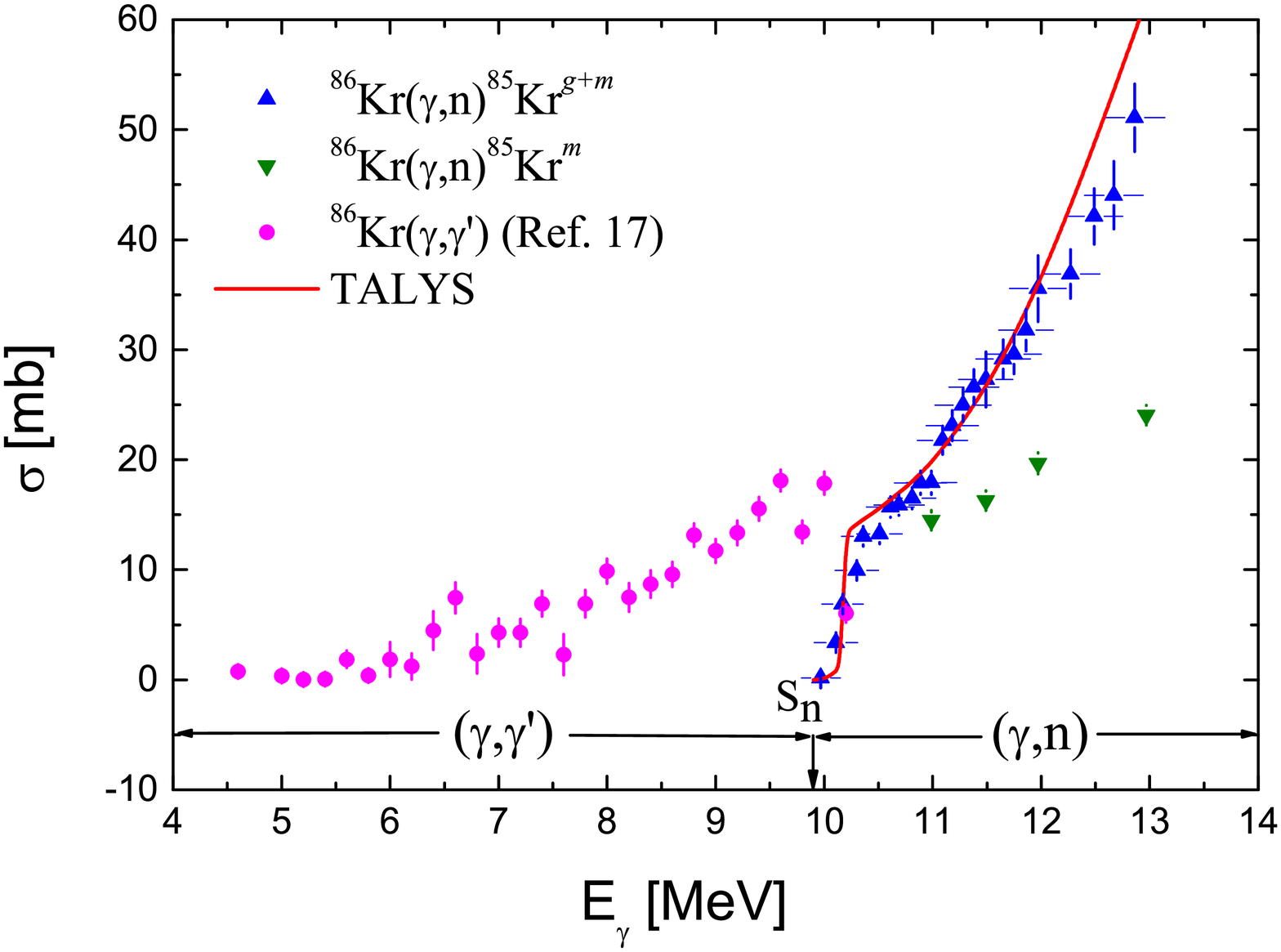}
\caption{\label{fig-results1} (Color online) The blue point-up triangles represent the total cross section of the $^{86}$Kr($\gamma,n$)$^{85}$Kr$^{g+m}$ reaction while the green point-down triangles denote the $^{86}$Kr($\gamma$,n)$^{85}$Kr$^m$ cross section, both from the present measurements. The calculated cross section for the $^{86}$Kr($\gamma,n$)$^{85}$Kr$^{g+m}$ reaction, using the TALYS code, is represented by the red line. The magenta points denote the total photoabsorption cross section of $^{86}$Kr($\gamma,\gamma^\prime$) from Ref. \cite{Sch13}. 
}
\end{figure}

\begin{figure}
\includegraphics[angle=0,scale=.30,clip=true]{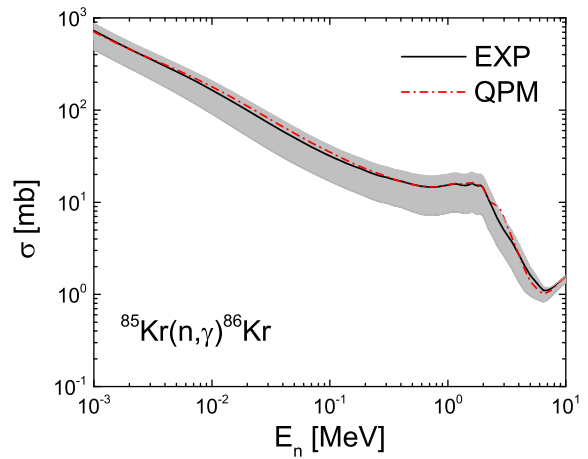}
\caption{\label{fig-results2} (Color online) Cross section of $^{85}$Kr$^g$$(n,\gamma)$$^{86}$Kr calculated with TALYS using experimental dipole (in black) and three-phonon quasiparticle-phonon model (QPM) strength functions (in red) from Ref. \cite{Sch13}. The predicted uncertainties (shaded area) are derived from the experimental errors of the dipole strength function and from variations in the nuclear level density parameters.}
\end{figure}

The cross-section values from the present measurements are plotted in Fig.~\ref{fig-results1}. The uncertainty on the flux determination is caused by the uncertainties of the cross-section values of the $^{197}$Au($\gamma,n$) monitor reaction \cite{Vog02}, and the neutron detector efficiency. The uncertainties were added in quadrature to estimate the total uncertainty on the cross-section values. It should be noted that the statistical uncertainties from the $^{3}$He proportional counter or the $^{197}$Au monitor reaction were less than 1\% at each incident beam energy. Hence, the total uncertainties of the cross-section values were dominated by the efficiency of the $^3$He proportional counter ($\sim$ 3\%) and the flux estimation using the $^{197}$Au($\gamma,n$) reaction ($\sim$ 5\%).\\ 

Similar to the analysis of previous ($\gamma$,n) measurements of interest to the $s$-process \cite{Son03, Has08, Mak09, Moh04}, we carried out statistical model calculations using the TALYS-1.43 code \cite{Kon08,Gor08b}. The results of these calculations based on the Hartree-Fock-Bogolyubov (HFB) + quasiparticle-random-phase approximation (QRPA) $\gamma$-ray strength \cite{Gor04b}, shifted by -0.5 MeV, are also plotted in Fig.~\ref{fig-results1} for the $^{86}$Kr$(\gamma,n)$$^{85}$Kr$^{g+m}$ reaction. \\ 

As shown in Fig.~\ref{fig-results1}, the $\gamma$-ray strength function is now fully constrained at all photon energies relevant to the $s$-process by our experimental data, below and above the neutron separation energy. The experimental $\gamma$-ray strength function, obtained by both the ($\gamma,n$) and the ($\gamma,\gamma^\prime$) measurements, was directly included in the TALYS code, in a tabulated and interpolated form, for calculating the neutron capture cross section of $^{85}$Kr. The transitional region around the neutron threshold has been described by the QRPA model, which is shown in Fig.~\ref{fig-results1} to reproduce the experimental data fairly well. The result of this calculation is presented in Fig.~\ref{fig-results2}.  The estimated uncertainties (shaded area) are dominated by the experimental errors of the $\gamma$-ray strength function (see Fig.~\ref{fig-results1}) and the assumption for the nuclear level density. For the latter quantity, the ``Hartree-Fock-Bogolyubov plus combinatorial model" \cite{Gor08} was adopted in calculating the recommended cross section, while the upper and lower cross section limits were obtained on the basis of the phenomenological models described in Ref.~\cite{Kon08b}. Note that the de-excitation strength function is assumed to be identical to the photoabsorption strength, i.e no temperature dependence is included in the theoretical strength function. The corresponding uncertainty at these low temperatures are believed to remain low with respect to those associated with the experimental errors on the ($\gamma,\gamma^\prime$) strength and the nuclear level densities. Furthermore, a dipole strength function from  three-phonon QPM calculations + QRPA, which has successfully described the fragmentation pattern of the E1 strength below the neutron threshold of $^{86}$Kr and the related pygmy dipole resonance \cite{Sch13}, was implemented in the TALYS code. These results are also shown in Fig.~\ref{fig-results2} and are found to be in very good agreement with the experimental data on the one hand and the HFB+combinatorial results on the other hand. The agreement confirms the predictive power of sophisticated many-body theoretical method like the QPM for exploratory investigations of n-capture reaction rates in hitherto experimentally inaccessible mass regions. At stellar temperature of $kT=30$ keV our new Maxwellian-averaged cross section amounts to a value of $83^{+23}_{-38}$ mb. This value is about 50\% higher than the result of Ref.~\cite{bao00} quoted above. Furthermore, the uncertainty is improved by a factor of about 3 to $\approx$50\%. \\

In order to study the impact of our new $^{85}$Kr($n,\gamma$)$^{86}$Kr reaction rate on the interpretation of the high $^{86}$Kr/$^{82}$Kr ratios measured in large stardust SiC grains, we simulate the $s$-process in AGB stars by employing stellar models computed previously using the Stromlo stellar structure code \cite{Kar10a,Kar10b}. The temperatures, densities, and convective velocities extracted from each model were input to a post-processing reaction network code that included 320 nuclides from hydrogen to bismuth linked by 2,336 nuclear reactions. Strong and weak interaction rates were adopted from the May 2012 version of the JINA reaclib \cite{Jin12}, except for the $^{85}$Kr($n,\gamma$)$^{86}$Kr reaction (see below). The reaction network was solved numerically together with convective mixing. \\ 

We considered stellar models of 1.25 M$_{\odot}$ and 1.8 M$_{\odot}$ with a metallicity of Z=0.01 \cite{Kar10b}, 3 M$_{\odot}$ with Z=0.02 \cite{Kar10a}, and 3 M$_{\odot}$ with Z=0.01 \cite{Shi13}. In all of these models dredge-up carries enough carbon to the envelope to ensure a higher carbon abundance relative to oxygen, which is a necessary condition for the formation of carbon-rich dust such as SiC. Models of higher mass than considered here suffer proton captures at the base of the convective envelope, thereby destroying carbon and resulting in a higher oxygen abundance relative to carbon. Two neutron sources have been identified to operate inside AGB stars \cite{Gal98}. The $^{22}$Ne$(\alpha,n)$$^{25}$Mg reaction is activated under convective conditions during thermal pulses when the temperature reaches above $\sim$ 300 MK, giving rise to high neutron densities ($\sim$10$^{13}$ n/cm$^3$). However, in low-mass AGB stars ($<$ 4 M$_{\odot}$), the temperature in the thermal pulse barely reaches 300 MK, the $^{22}$Ne source is only very marginally activated, and another neutron source, the $^{13}$C$(\alpha,n)$$^{16}$O reaction, is at work at lower temperatures ($\sim$ 90 MK). To produce $^{13}$C it is assumed in the models that a small amount of protons is mixed from the envelope into the helium-rich shell during each dredge-up episode. These protons react with the abundant $^{12}$C via $^{12}$C($p,\gamma$)$^{13}$N($\beta^+$)$^{13}$C to produce a thin region rich in $^{13}$C, the $^{13}$C “pocket”, where $^{13}$C burns in-between thermal pulses under radiative conditions producing relatively low neutron densities ($\sim$ 10$^8$ n/cm$^3$). Our models include the $^{13}$C pocket by artificially introducing, at the end of each dredge-up episode, a proton profile that decreases exponentially over a mass of 0.002 M$_{\odot}$ just below the base of the convective envelope \cite{Lug04,Gor00}. In all the models most of the $^{13}$C in the pocket is consumed before the onset of a subsequent thermal pulse, except in the case of the 1.25 M$_{\odot}$ model, where the temperature is too low for this to occur and a significant amount of $^{13}$C is left behind for burning during the following thermal pulse (see also \cite{Cri09,Cri11,Lug12}). In these conditions the $^{13}$C neutron source burns at higher temperatures ($\sim$ 200 MK rather than $\sim$ 90 MK), the burning timescale is shorter, the neutron density is higher, and the $^{85}$Kr branching point is more activated.

\begin{figure}
\includegraphics[angle=0,scale=.35]{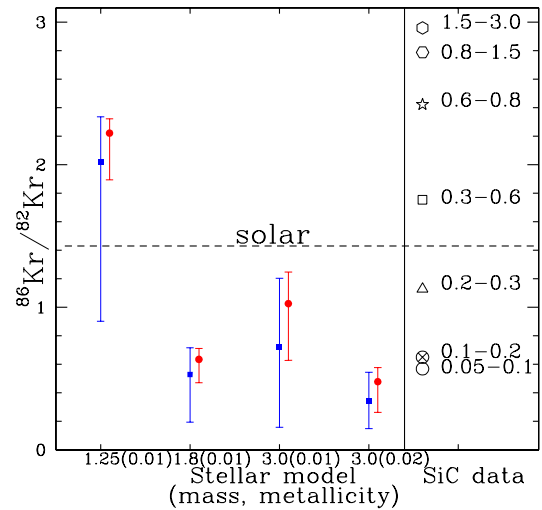}
\caption{\label{fig-grains}
(Color online) Number abundance ratio $^{86}$Kr/$^{82}$Kr from different stellar models of AGB stars. Results shown as red circles and blue squares are obtained using the present and previous $^{85}$Kr$^g$($n,\gamma$)$^{86}$Kr reaction rate, respectively. Values observed in stardust SiC grains are shown on the right-hand side; the numbers next to the symbols denote the size of the grains in units of $\mu$m. The solar ratio is shown as horizontal dashed line.}
\end{figure}

Observed and predicted $^{86}$Kr/$^{82}$Kr isotopic ratios are displayed in Fig.~\ref{fig-grains}. The values measured in stardust SiC grains of different sizes are shown as black open symbols and display a clear increase of the $^{86}$Kr/$^{82}$Kr ratio with increasing grain size, indicative of a different implantation energy of the $^{86}$Kr-rich component. The predicted ratios are derived from the composition of the helium-rich shell at the end of the evolution for each stellar model considered in the present work. The red and blue data points correspond to the values predicted using the present and previous $^{85}$Kr$^g$$(n,\gamma)$$^{86}$Kr reaction rate, respectively. It can be seen that the $^{86}$Kr/$^{82}$Kr ratios obtained using our new $^{85}$Kr$^g$$(n,\gamma)$$^{86}$Kr rate are on average 40\% higher compared to the results derived with the previous rate \cite{bao00}. As already pointed out, our new rate is based on experimental information for the $^{85}$Kr+n system. Varying the $^{85}$Kr$^g$$(n,\gamma)$$^{86}$Kr rate within the present uncertainties changes the predicted $^{86}$Kr/$^{82}$Kr ratio by a factor of $\sim$ 2, while a change by a factor $\sim$ 7 is obtained with the previous reaction rate. Most of the models presented here cannot match the high $^{86}$Kr/$^{82}$Kr ratios observed in the large SiC stardust grains because the $^{22}$Ne$(\alpha,n)$$^{25}$Mg neutron source operates only marginally in these models and, consequently, the $^{85}$Kr branching point is only weakly activated. The exception is the 1.25 M$_{\odot}$ model, where most of the $^{13}$C neutron source burns convectively resulting is in a better agreement with the higher observed ratios in large-size SiC stardust grains. \\

In conclusion, our first experimentally-based determination of the $^{85}$Kr($n,\gamma$)$^{86}$Kr cross section {\it{sets the necessary, essential premise}} to proceed in the investigation of the origin of the high $^{86}$Kr/$^{82}$Kr ratio observed in large SiC grains. We can conclude that our 1.25 M$_{\odot}$ AGB model points to a possible explanation for the $^{86}$Kr/$^{82}$Kr composition observed in large-size SiC stardust grains. Other plausible solutions, for
example, involving proton-ingestion episodes occurring during the post-AGB phase \cite{Her11} can now also be investigated, thanks to the new cross section presented here. \\

This work is supported by the US DOE grants DE-FG52-09NA29448, DE-PS52-08NA28920, DE-FG02-97ER41033, DE-FG02-97ER41041, and DE-FG02-97ER41042. ML is an ARC Future fellow and Monash Fellow. The authors would like to thank Dr. T. Kawano for valuable discussions. We thank Amanda Karakas for providing the stellar structure of the AGB models. \\

\bibliography{86Kr_arxiv}
\end{document}